# Selective intracellular delivery and intracellular recordings combined on MEA biosensors

Andrea Cerea[1,*], Valeria Caprettini[1,*], Giulia Bruno[1,2], Laura Lovato[1], Giovanni Melle[1,2], Francesco Tantussi[1], Rosario Capozza[1], Fabio Moia[1], Michele Dipalo[1,*], Francesco De Angelis[1,*]

[1]Istituto Italiano di Tecnologia, 16163 Genova, Italy
[2]DIBRIS, Università degli Studi di Genova, 16145 Genova, Italy

Email: francesco.deangelis@iit.it, michele.dipalo@iit.it

**Abstract**

Biological studies on *in vitro* cell cultures are of fundamental importance to investigate cells response to external stimuli, such as new drugs for treatment of specific pathologies, or to study communication between electrogenic cells. Although three-dimensional (3D) nanostructures brought tremendous improvements on biosensors used for various biological *in vitro* studies, including drug delivery and electrical recording, there is still a lack of multifunctional capabilities that could help gaining deeper insights in several bio-related research fields. In this work, the electrical recording of large cell ensembles and the intracellular delivery of few selected cells are combined on the same device by integrating microfluidics channels on the bottom of a multi-electrode array decorated with 3D hollow nanostructures. The novel platform allows to record intracellular-like action potentials from large ensembles of cardiomyocytes derived from human Induced Pluripotent Stem Cells (hiPSC) and from the HL-1 line, while different molecules are selectively delivered into single/few targeted cells. The proposed approach shows high potential for enabling new comprehensive studies that can relate drug effects to network level cell communication processes.

## Introduction

Intracellular delivery is a fundamental method to study diseases and to investigate biological processes that affect cell functioning and communication with the extracellular environment. Plasmids, nanoparticles, fluorescent dyes or any bio-molecule need to be injected into cells to study their effects on physiology[1]. However, their introduction in the intracellular compartment is often challenging due to the cellular membrane that acts as barrier. Cytosolic delivery relies on established methods to transfer reagents across the cell membrane, such as viral vectors[2] for single-cell gene transfection and endocytosis of specific carriers (lipofectamine, cationic polymers). However, these methods are often hampered by lack of specificity, low efficiency, high expenses and even toxicity to the cell itself, as well as for specifically inhibiting signal transduction pathways that often involve small peptide inhibitors of protein kinases that are not cell-permeant. These drawbacks have led to the development of more direct approaches to physically breach the cell membrane, among which micropipetting and electroporation are the most common[3,4]. Furthermore, recent advances in nanofabrication techniques, together with a deepened understanding of the cell-substrate interaction[5,6], have enabled the realization of nanostructured platforms for improved intracellular delivery[7,8], providing selectivity, lower invasiveness and better localization of the delivery process. As we have recently shown, using high aspect ratio 3D hollow gold nanoelectrodes in combination with electroporation or optoporation, it is possible to open transient nanopores into the cell membrane with controlled spatial localization and without compromising the cell/nanoelectrode coupling, allowing for intracellular delivery controlled in space and time[9,10].

For the investigation of neurons and cardiomyocytes, which are among the most important cells in regard to drug development and cell biology, the monitoring of their electrical activity is fundamental for their proper characterization, especially in large cultures where collective behavior can influence the functioning of single cells[11]. Consequently, biological experiments in which molecules or drugs are delivered to these cells should be implemented also in their recording capabilities to gain comprehensive insights into the delivery effects. However, the fabrication complexity of nano-fluidic nanostructures prevented so far the realization of devices that could integrate intracellular delivery and electrical recording on the same platform. In fact, with biosensors for electrical recording of electrogenic cell cultures, such as multi-electrode arrays (MEAs) and CMOS-MEAs[12–14], the common drug delivery methods are mostly performed on entire cellular networks, which are globally affected by the treatment under investigation. Deeper and more crucial understandings could be gained by investigating the firing activity of large cell populations in which only single or few identified cells have been addressed by intracellular delivery. For example, pathologies may be induced to only few cells in order to study the reaction of the whole culture, simulating thus the early stage development of several diseases, or alternatively, drugs may be administered on partial regions of a cell culture to evaluate the electrical response in the case of inhomogeneous drug distribution[15]. In the case of cardiomyocytes, the signal propagation over the culture could reveal unexpected behaviors of the syncytia between delivered and unaffected cells. In recent years, several approaches have been presented to implement microfluidics on MEA through PDMS (Polydimethylsiloxane) or polymer channels placed onto the surface of the arrays[16–18]. These strategies provided ways, for example, to configure different culture conformations by confining cell growth into compartments. However, they do not reach the spatial resolution required to address single cells for drug delivery nor do they provide a nanoscale interface for intracellular access. In an attempt to fill this technological gap, here we present a novel multifunctional MEA biosensor that combines network level electrical recording with single cell intracellular delivery, realizing a complete and ready-to-use platform for extra-/intracellular recordings of electrogenic cells (cardiomyocytes) and simultaneous molecule delivery in a non-invasive and selective way (figure 1**a**). The proposed system consists of a MEA design fabricated on a very thin silicon nitride membrane and decorated with three-dimensional (3D) hollow nanostructures. In this configuration, the MEA layout allows recording of large ensembles of cells and addressing of few to single cells with high spatio-temporal control, whereas the hollow nanostructures enable controlled delivery of reagents in specific locations of the cell culture.

## Results and discussion

**The biosensor concept**

The novel platform is built on the well-established concept of the passive MEA device, which consists of a matrix of planar gold electrodes and represents the gold standard for the electrical characterization of large cell networks. In our approach, the typical elements of the MEA structure (electrodes, feed lines, passivation) are fabricated on a thin silicon nitride layer (500 nm) deposited on a silicon wafer (see a representative sketch in figure 1**a** and the MEA photograph in figure 2**c**). On the backside of the MEA we selectively remove the silicon beneath the $Si_3N_4$ layer, forming two or more thin nitride membranes (see figure 2**d**, **f** and figure **S3**) that run underneath the gold electrodes and act as microfluidic channels (micro-channels). Each MEA electrode is decorated with 3D gold nanotubes that are through-hole (see figure 2**e**, **g**), thus connecting the two sides of the thin nitride membrane. In essence, exploiting the favourable geometric characteristics of 3D hollow metallic nanopillars, we can take advantage of electroporation to get direct access to the cell interior with high spatial localization and perform multiple functionalities at the same time. On the one hand, they act as nano-fluidic channels for the flow of molecules or drugs from the underlying channels, providing a way to deliver them locally only to the cells lying on the nanotubes. On the other hand, the gold nanotubes are connected to the planar MEA electrode and offer electrical recording with major improvements over signal quality and spatio-temporal resolution for the recording of large population of cells[19,20], increasing the cell-electrode seal resistance and thus reducing parasitic signal losses.

For our experiments, we used a MEA array of 24 electrodes placed in a 4×6 matrix and with an inter-electrode pitch of 400 µm, resulting in an active area of 2×1.2 mm$^2$. Beneath the active area, we fabricated different microfluidic configurations with 2, 4 and 6 independent channels (see figure 1**b-d**)**,** each interfacing several electrodes decorated with 3D hollow nanoelectrodes (figure 2**g**). These microfluidic configurations could be fabricated with the same techniques, with no adjustment required for the fabrication of the samples presenting the higher number of channels. In order to reduce the complexity of the tubing during electrophysiological experiments, we accumulated statistics using the devices with two microfluidic channels for cell culture, electrical recording and fluorescence imaging.

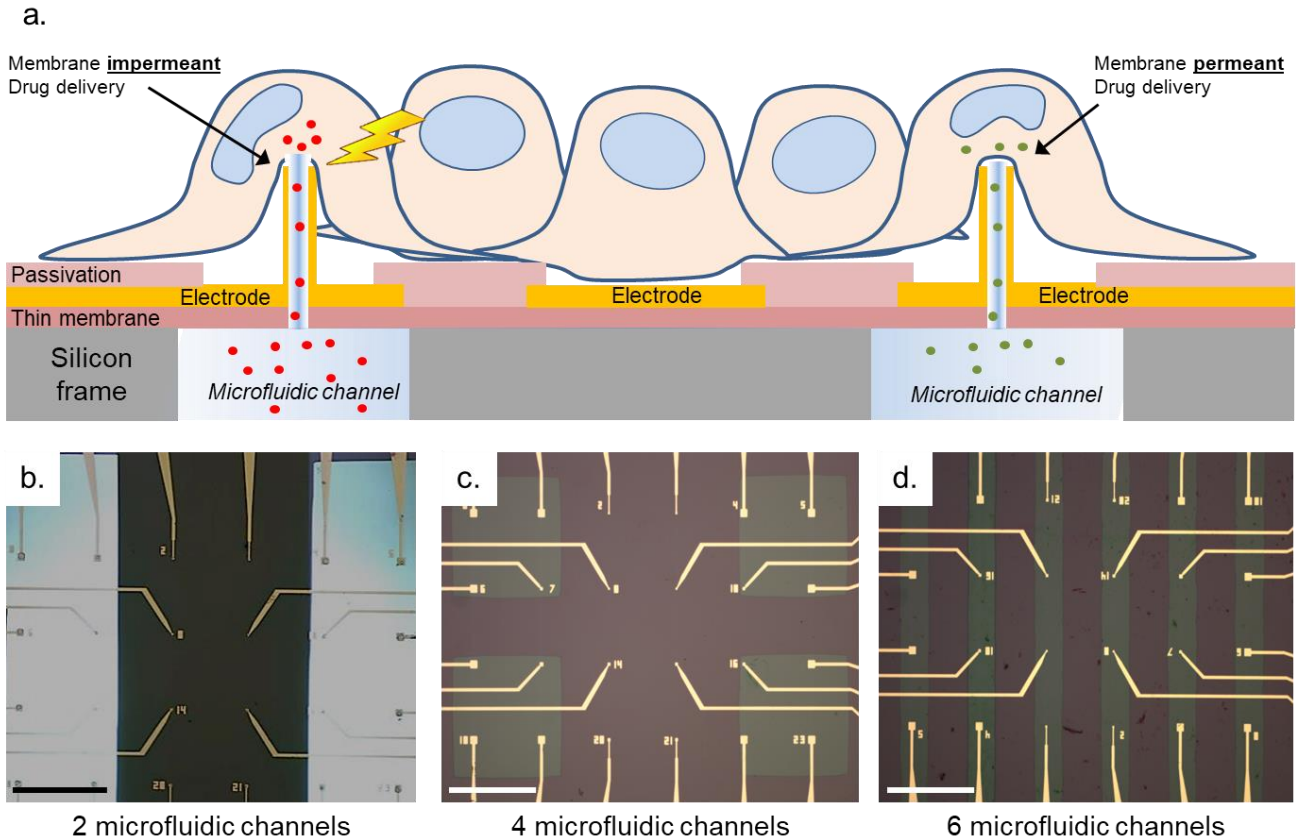

**Figure 1**. The biosensor concept. **a**) Cross-sectional sketch of the MF-MEA biosensor enabling multiple functionalities at the same time. **b**)-**d**) Optical images of various microfluidic configurations with, respectively, 2, 4 and 6 independent micro-channels. The different colours of the substrate refer to the silicon nitride membranes (brighter) and bulk silicon frame (darker). The scale bars are 500 μm.

**Electrophysiological recording of cardiomyocyte spontaneous beating and electroporation**

The microfluidic-MEAs (MF-MEAs) were packaged using a custom-designed PCB board (figure 2**a**) that allows for the tubing access from underneath (figure 2**b**). Even with the integrated multiple capabilities, the proposed platform remains still compatible with standard MEA acquisition systems such as those from Multichannel Systems GmbH.

Extracellular and intracellular-like recordings were performed on cardiomyocytes derived from human Induced Pluripotent Stem Cells (hiPSC) and on HL-1 cells derived from rat cardiomyocytes, using our nanostructured electrodes both as sensing elements and for electroporation. The hiPSC derived cardiomyocytes represent a promising future benchmark for drug screening experiments on cardiac cells and are being investigated comprehensively by the scientific community[21–23]. On the other hand, the cells of the HL-1 line, having been characterized extensively in literature[20,24,25], represent a well-established experimental bench to test the recording performance of newly designed MEAs. Experiments were executed outside the incubator keeping the temperature at 37 °C by means of a Peltier cell attached to a properly designed aluminum cage surrounding the culture well. In order to maintain sterility without affecting $CO_2$ exchange, a PDMS thin membrane was attached onto the glass ring prior to experiments. The compact MEA biosensor was then mounted on a custom-designed amplifier (MEA acquisition system) acquiring data from 24 electrodes at a 10 kHz sampling rate. Electrical recordings with the MF-MEAs were obtained in 6 cultures and with 10 different devices (4 cultures and 6 devices with HL-1 cells; 2 cultures and 4 devices with hiPSC cardiomyocytes). The recordings were performed after approximately 4-5 days-in-vitro (DIV) to let the culture reach confluence, which is necessary to obtain a spontaneous synchronized firing. Figure 2**h** shows examples of hiPSC human cardiomyocyte spontaneous electrophysiological activity from one of the 24 channels of a MF-MEA at 4 DIV. The signal-to-noise ratio (SNR) of the extracellular spikes is considerably high, allowing for the full characterization of the cardiomyocyte culture in terms of beating frequency and propagation patters. Hence, the presence of the microfluidic channels below the device does not degrade the electrical recording performance of the MF-MEAs. To achieve intracellular recording, we use a soft electroporation protocol that exploits the 3D nature of the hollow nanoelectrodes to porate the cellular membrane[10]. A train of voltage pulses (consisting of 200 consecutive symmetric square pulses, each with a 50 ms period, a 100 µs duration and 2.5 V amplitude) were applied through single nanostructured electrodes, without affecting the rest of the cell culture. Recordings were resumed less than 30 s after electroporation and showed (see figure 2**i**) a clear change of both spike duration and waveform, together with polarity inversion that is typical of intracellular-like signals. Most importantly, the duration of the spike increases from few milliseconds to more than 50 ms after the switch to the intracellular-like configuration (figure **S4**), allowing for the proper characterization of drugs on cardiac cells. In fact, the effects of many drugs on the heart are mainly determined by changes of depolarizing and repolarizing currents, which are quantified by the spike duration at different amplitude levels. Interestingly, the post-electroporation signals maintained intracellular characteristics over several minutes with a slow reduction in amplitude over time, presumably due to resealing of transient membrane pores. The spontaneous activity resumes extracellular features after 10 – 15 minutes as already observed in literature for similar cells after electroporation[20]. In supporting information we report additional electrical measurements performed on hiPSC cardiomyocytes (figure **S4**) and HL-1 cells (figure **S5**).

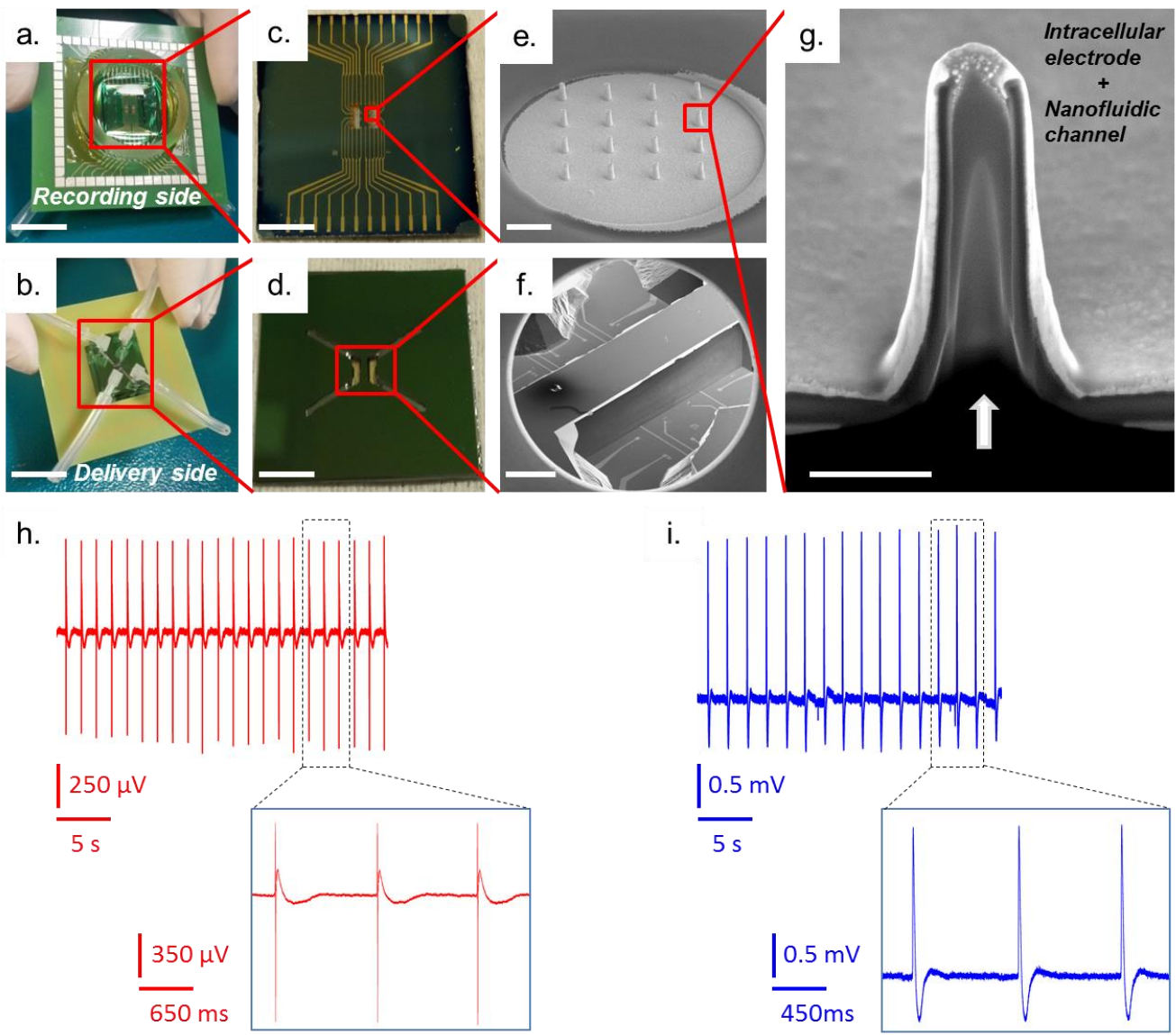

**Figure 2**. MEA device and electrical recordings. **a**)-**b**) Respectively top and bottom view of the packaged MF-MEA. The scale bar is 0.5 mm. **c**)-**d**) Top and bottom view showing, respectively, the MEA, feedlines and contact pads and the independent microfluidic channels underneath. **e**) Scanning Electron Microscopy (SEM) image of a MEA electrode decorated with gold-coated hollow nanostructures and surrounded by a SU-8 passivation layer. The scale bar is 5 μm. **f**) SEM image of the microfluidic channels selectively etched underneath the active area of the MF-MEA. The MEA electrodes at the other side are visible through the thin silicon nitride membranes. The scale bar is 500 μm. **g**) SEM image of a Focused Ion Beam (FIB) cross-sectioned nanoelectrode connecting the microfluidic compartment with the culture well. The scale bar is 500 nm. **h**) Spontaneous electrophysiological spiking activity of hiPSC cardiomyocytes from one of the 24 channels of a MF-MEA at 4 DIVs and detail of three

consecutive spikes. **i**) Intracellular-like activity of hiPSC cardiomyocytes acquired on the same electrode after electroporation and detail of three consecutive spikes.

**Intracellular delivery**

In order to test the effectiveness of our multifunctional platform, we performed selective delivery of molecules through the 3D nanoelectrodes during the recording sessions. The hollow nanoelectrode design enables to address only few selected cells, without affecting the electrical activity of the whole network. As a first proof-of-concept of our technology, calcein-AM, a membrane permeant molecule producing a green fluorescence at λ=515 nm, was delivered at a concentration of 200 µM in a Phosphate Buffer Solution (PBS) 1X to the cells (electroporated or not) interfacing the 3D nanoelectrodes on the $Si_3N_4$ membrane. The desired solution was previously incubated at 37 °C, and was then injected with our pumping system into the two isolated microchannels, allowing the fluorophores to diffuse through the nanoelectrode just in small portions of the biosensor surface. Figure 3**a** shows the green fluorescence image acquired with an optical microscope (air objective 5×) while the whole cell culture was live and electrically active. The fluorescence image were taken 1 hour after the addition of calcein-AM in the microfluidic channels below the MEA. This time was needed to allow the calcein-AM to diffuse into the 3D hollow nanoelectrodes and to access the cellular membrane. Before imaging, the calcein-AM solution was then substituted by PBS in order to rinse both the tubing and the microchannels, removing calcein-AM residuals. As it can be seen, calcein-AM is selectively delivered only to the cells that are tightly sealed on top of the nanostructured electrodes. The delivery selectivity reaches almost 100% thanks to the tight cell adhesion on the hollow nanopillars, which prevents the diffusion of the dye molecule into the rest of the cell culture.

In other experiments, we injected into the microfluidic channels a saline solution with Propidium Iodide (PrhD-1), a non-permeant molecule which binds nucleic acids and emits a bright red fluorescence at λ=635 nm, with a concentration of 1 mM in PBS 1X. Electroporation pulses were applied through the 3D nanoelectrodes of the MEA electrodes interfacing the microfluidic channels in order to open transient pores in the cellular membrane and to deliver the PrhD-1 inside the cells. As it can be seen in figure 3**b** and 3**c**, PrhD-1 is delivered only to the cardiomyocytes laying on the electrodes and hence subjected to electroporation, thus providing a clear evidence of the high selectivity and spatial control achieved by our multifunctional platform. We also investigated the possibility to deliver non-permeant and permeant molecules through the same microfluidic channels. Figure 3**d**, **e**, **f** shows bright-field and fluorescence images of four MEA electrodes (A, B, C and D) after we delivered both calcein-AM and PrhD-1 and applied selective electroporation only to electrodes B and D. We observe the selective delivery of calcein-AM on the cells laying on the four electrodes (figure 3**e**), whereas we note that PrhD-1 is delivered only to the electroporated cells laying on the electrodes B and D (figure 3**f**). The fluorescence images were acquired simultaneously with the electrical recording of the same cells, which indeed showed polarity inversion of the recorded signals due to the intracellular action potential contribution (insets in figure 3**e**, **f**). The same experiments were performed on different sets of electrodes, finding good results in terms of reproducibility and intracellular selectivity (see figure S6). The cells were fixated after recording and delivery in order to acquire high resolution images of fluorescence emission with a confocal microscope. Figure 3**g** shows the detail of one electrode on which we have delivered calcein-AM and PrhD-1 and applied electroporation. We can observe the red staining of the nuclei due to PrhD-1 and the surrounding green fluorescence due to the calcein-AM staining the cell cytoskeleton.

Further, we studied the effects of culturing the cells over a heterogeneous sample consisting of areas on bulky silicon and other regions on thin nitride membrane with PBS underneath, so as to evaluate possible divergences in cell proliferation. HL-1 cells were cultured on the microfluidic MEAs and stained with DAPI, a blue staining of the nuclei, after 4 DIV. In this case, the staining with DAPI was performed on the complete culture from the glass culture well, rather than selectively from the microfluidic channels. The images reveal a homogeneous cell growth over the whole MEA surface, with no distinction between the free-standing nitride membranes and the bulk silicon part (figure 3**h**). This ensures that the particular employed device configuration does not influence the cell culture growth and development. Moreover, the cell density estimate reported in figure S7 yields similar values on different areas (bulk silicon or nitride membranes) of the same device. Figure 3**i** shows a detail of four electrodes after delivery of calcein-AM through the 3D nanoelectrodes and staining with DAPI from the cell culture well.

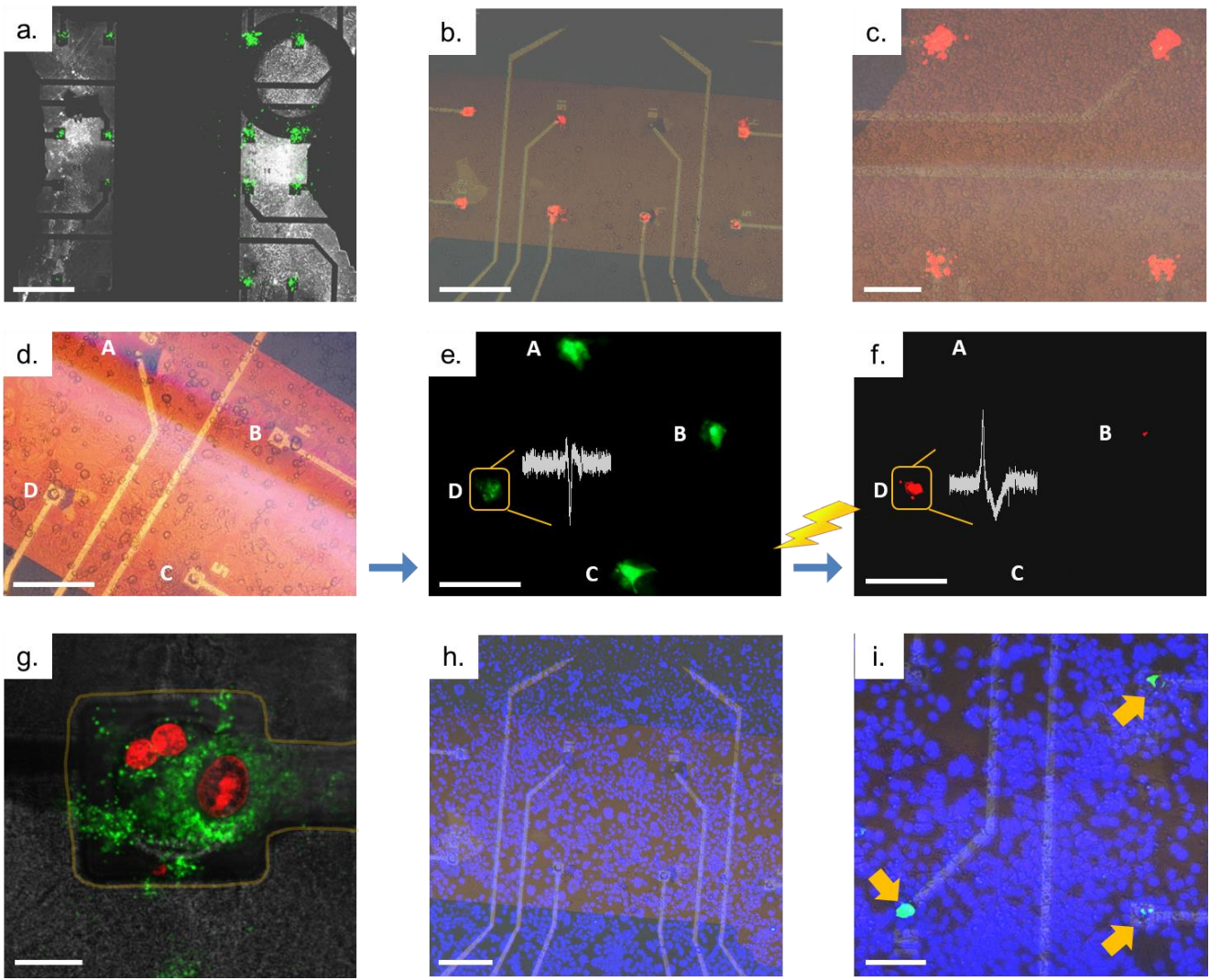

**Figure 3**. Selective intracellular drug delivery on MF-MEA. **a**) Fluorescence image of intracellular delivery of calcein-AM on all the MEA nanostructured electrodes. The scale bar is 500 μm. **b**) Fluorescence image of intracellular delivery of PrhD-1 into selectively electroporated cardiomyocytes. The scale bar is 300 μm. **c**) Fluorescence image of intracellular delivery of PrhD-1 into electroporated cardiomyocytes onto four membrane electrodes. The scale bar is 100 μm. **d**) Bright-field optical image of four electrodes of a MF-MEA. The scale bar is 200 μm. **e**) The same 4 electrodes observed with a green fluorescence filter, showing calcein-AM delivery to the cells onto the 4 electrodes. The scale bar is 200 μm. **f**) The same four electrodes observed with a red fluorescence filter, showing PrhD-1 delivery to the cells onto the 2 electrodes where electroporation was performed. The scale bar is 200 μm. **g**) Confocal image of a nanostructured electrode where both calcein-AM and, after electroporation, PrhD-1 have been delivered to the attached cells. The scale bar is 10 μm. **h**) Fluorescence image of cardiomyocytes cultured on MF-MEA and stained with DAPI after 4 DIV. The scale bar is 200 μm. **i**) Fluorescence image of four electrodes after delivery of calcein-AM through the 3D nanoelectrodes and staining with DAPI from the cell culture well. The scale bar is 100 μm.

## Experimental

**Fabrication of nitride membrane and independent microfluidic channels**

In order to obtain a compact and flow-through design, nanostructured MEAs were realized on thin silicon nitride ($Si_3N_4$) membranes. Membranes and micro-channels were fabricated on a 525 µm thick p-doped (100) Si wafer with 500 nm of LPCVD super low stress $Si_3N_4$ on both sides (see figure **S1** for fabrication protocol). A 400 nm thick mask layer of Cr was sputter coated on one of the two wafer sides and a thin layer (≈ 2 µm thick) of a common photoresist (MICROPOSIT S1813) was spin-coated on top. Conventional optical UV lithography (SÜSS MicroTec Mask Aligner) was employed to expose the desired areas (see figure **S1**). After exposure and 60 seconds development in (MICROPOSIT MF-319 DEVELOPER), Cr was removed by chemical wet etching in a ceric ammonium nitrate-based mixture (Sigma-Aldrich). The remaining unexposed resist was then dissolved in acetone, leaving Cr as an etch mask for the Reactive Ion Etching (RIE, SENTECH) process on the $Si_3N_4$ underneath. A mixture of gases, $CHF_3$ and $O_2$ (at 70 and 10 SCCM respectively, where SCCM is the standard cubic centimeter per minute) were used to selectively etch the 500 nm thick $Si_3N_4$; the process parameters of pressure, temperature and power were held at 1 Pa, 20 °C and 30 W respectively. As a last step, the Cr mask was chemically removed and Si was wet etched in a solution of 30% wt. potassium hydroxide (KOH) in deionized water (DI). Given the zero etch-rate of $Si_3N_4$ under KOH exposure, the wet etching process stops when the nitride is reached at the other side of the wafer in correspondence with the membrane apertures, while in the micro-channels stops at the meeting point between the etching planes; the final depth of the channels can be varied accordingly to the geometrical parameters of the etching area. The present microfluidic design can be easily rescaled from 2 to 6 independent channels configurations (see figure 1**b-d** and figure **S3**), therefore allowing multiplex applications on the same platform.

**Microfluidic-MEA**

MEAs decorated with 3D nano-electrodes were fabricated using the technologies developed in other works (Dipalo M. et al., Nanoscale 2015, De Angelis F. et al., Nano Letters 2013). Briefly, a thin layer (≈ 2 µm thick) of S1813 was spin-coated on the $Si_3N_4$ membrane (see figure **S2** for fabrication protocol). After UV exposure and 60 seconds development in MF-319, 24 Ti/Au electrodes, conductive tracks and contact pads were fabricated through electron beam evaporation in a high vacuum chamber (base pressure $10^{-7}$ mbar) with a 0.3 Å/s deposition rate. The unexposed resist and the metal on top of it were removed through a conventional lift-off process in hot Remover PG (MicroChem). The substrate was then rinsed out in isopropanol (IPA) and dried with nitrogen. $O_2$ plasma ashing at 100 W for 5 minutes was used to remove residual S1813 resist and other organic residues. 24 electrode contacts of varying dimensions (from 100 to 900 µm$^2$) were defined by a passivation of SU-8 that covered the rest of the device (see figure 2**e**). Gold coated 3D nanoelectrodes (figure 2**e**) were fabricated on the planar electrodes by focused-ion-beam (FIB) lithography (setting the exposure dose to 27 nC/µm$^2$)$^{26}$ from the backside of the Si$_3$N$_4$ membrane. The obtained hollow nanotubes are characterized by a conical shape (figure 2**g**), with external diameter of about 600 nm and height of 1.3 µm. Due to the fabrication protocol, gold nanoelectrode arrays are realized only on those electrodes lying on the nitride membranes, with their number depending on the desired microfluidic configuration. In the following, we will only consider the case of two isolated membranes with 8 nanostructured electrodes each. The remaining 8 planar electrodes will be used as controls.

**MEA packaging**

The complete MEA with nanostructured electrodes was then mounted on a PCB (printed-circuit-board) and passivated with Polydimethylsiloxane (PDMS) that covers the electrical connections with the PCB. Due to its chemical characteristics, PDMS ensures non-toxicity, inertness and biocompatibility. A glass ring was attached on the MEA by means of PDMS to form a culture well (see figure 2**a**). After this, the open micro-channels on the bottom side of the MEA were sealed with a piece of polymerized PDMS. Tight bonding between the Si$_3$N$_4$ and the PDMS was ensured by chemical activation of both surfaces via an oxygen plasma treatment for 30 s at a pressure of 0.5 mbar, 25 SCCM of oxygen and 30 W power. In order to improve the bonding stability and efficiency, the sample was then baked in a hoven at 80 °C for 30 minutes. The PDMS bonding step assures the realization of two non-communicating chambers with independent inlet and outlet, thus enabling the selective delivery of reagents in different areas of the MEA. As a last step, using biopsy punches holes were drilled through the PDMS piece (in correspondence with the micro-channels underneath) and tubing was added to provide external connections with our pumping system (see figure 2**b**).

**Pumping system**

A commercial pressure-driven microfluidic flow control system (MFCS-EZ, FLUIGENT) was employed to selectively deliver different solutions in the microfluidic chambers underneath the nitride membranes. By connecting the system pressure outputs to airtight reservoirs, containing the desired reagents, we were able to precisely control the liquid flow into the microfluidic device. The flow-rate was adjusted to 100 µL/min to ensure flow stability and smoothness.

**Culture of HL-1 cardiomyocytes**

The MF-MEAs have been sterilized with 20 minutes-UV exposure in laminar flow hood and incubated overnight (o.n.) with Claycomb culture medium in order to saturate the PDMS porous matrix of the culture well. Furthermore, the micro-channels and tubing at the bottom of the device have been manually filled with Claycomb using a sterile syringe. The day after, the medium has been completely removed from the culture well and the substrates have been treated with poly-L-lysine 0,01% w/v (Sigma-Aldrich) for five minutes, washed extensively with water and let dry in sterile condition. Subsequently, HL1 cells have been seeded at the density of 35000/cm$^2$ and grown with Claycomb, supplemented with 10% fetal bovine serum, 100 µM norepinephrine, 300 µM ascorbic acid, 2 mM L-glutamine, penicillin and streptomycin (Sigma-Aldrich). The culture medium has been changed daily. When 100% confluence was reached, the electrical activity of cells has been recorded.

**Human induced Pluripotent Stem Cells-derived cardiomyocytes**

HiPSCs (Axiogenesis) have been thawed and pre-cultured in a cell culture flask and growth o.n. in Cor.4U complete medium (Axiogenesis) before seeding on samples. This procedure allows removal of dead cells prior to seeding and will result in better assay performance. The MF-MEAs have been sterilized with 20 minutes-UV exposure in laminar flow hood and incubated o.n. with Cor.4U complete medium in order to saturate the PDMS passivation of the electrical connections. After the incubation, samples were treated in order to promote the tight adhesion of cells. The substrates were coated with Geltrex ready-to-use solution (Thermo Fischer Scientific) and then incubated for 30 minutes at 37°C in a humidified environment. As a following step, the solution was totally removed and cells were rapidly plated without the coating was allowed to dry. HiPSCs-derived cardiomyocytes have been plated at the density of 70000/ cm$^2$ and grown with Cor.4U complete medium.

**Cell fixation**

Samples were fixed with PFA 4% in PBS. After a 30 minutes incubation at room temperature, samples have been extensively washed with PBS and treated with Triton 0.1% in PBS and stained with DAPI (Invitrogen) in order to mark the nuclear compartment of cells.

**Fluorescence imaging and confocal microscopy**

Confocal imaging of HL-1 cardiomyocytes was performed with a Leica TCS SP5 AOBS Tandem DM6000 upright microscope equipped with a 25× objective (Leica IRAPO, NA = 0.95). Data were subsequently analyzed by LASX V2.0 software (Leica Microsystems). Fluorescence imaging during electrical recording was performed with a Nikon FN-1 microscope using 5× and 20× air objectives and a 60× water-immersion objective.

# Conclusion

In this work, we presented a novel multifunctional biosensing platform with high level integration of microfluidics and recording capabilities. The approach combines various bio-nanotechnologies to perform selective intracellular delivery of molecules, extracellular action potential recording and intracellular recording by means of electroporation. We successfully performed intracellular recording of human iPSC derived cardiomyocytes and intracellular delivery of two different molecules in few selected cells, while the electrical activity of the whole cultures was monitored.

The peculiar properties of the 3D gold hollow nanostructures and of the multiple microfluidic channels will allow to inject molecules into selected cells and to measure the consequent response of the whole cell network. Moreover, different molecules may be injected in various and specific parts of the cell culture through independent microfluidic channels, so that the electrical response of the cell culture can be observed also in the presence of multiple drugs or molecules. The platform may provide new methodologies to study early stage pathologies and to evaluate toxicity of drugs, molecules and nanoparticles. In perspective, the same configuration may be used in the future also for performing intracellular sampling as already suggested by Y. Cao et al.[27], combining intracellular sampling and electrical recording of the same cells in a large network.

## Author contributions

A.C and V.C contributed equally to the work.

## Conflicts of interest

There are no conflicts to declare.

## Acknowledgements

The research leading to these results has received funding from the European Research Council under the European Union's Seventh Framework Programme (FP/2007-2013)/ERC Grant Agreement no. [616213], CoG: Neuro-Plasmonics.

# Supporting Information

# Selective intracellular delivery and intracellular recordings combined on MEA biosensors

Andrea Cerea[1,*], Valeria Caprettini[1,*], Giulia Bruno[1,2], Laura Lovato[1], Giovanni Melle[1,2], Francesco Tantussi[1], Rosario Capozza[1], Fabio Moia[1], Michele Dipalo[1,*], Francesco De Angelis[1,*]

[1]Istituto Italiano di Tecnologia, 16163 Genova, Italy
[2]DIBRIS, Università degli Studi di Genova, 16145 Genova, Italy

Email: francesco.deangelis@iit.it, michele.dipalo@iit.it

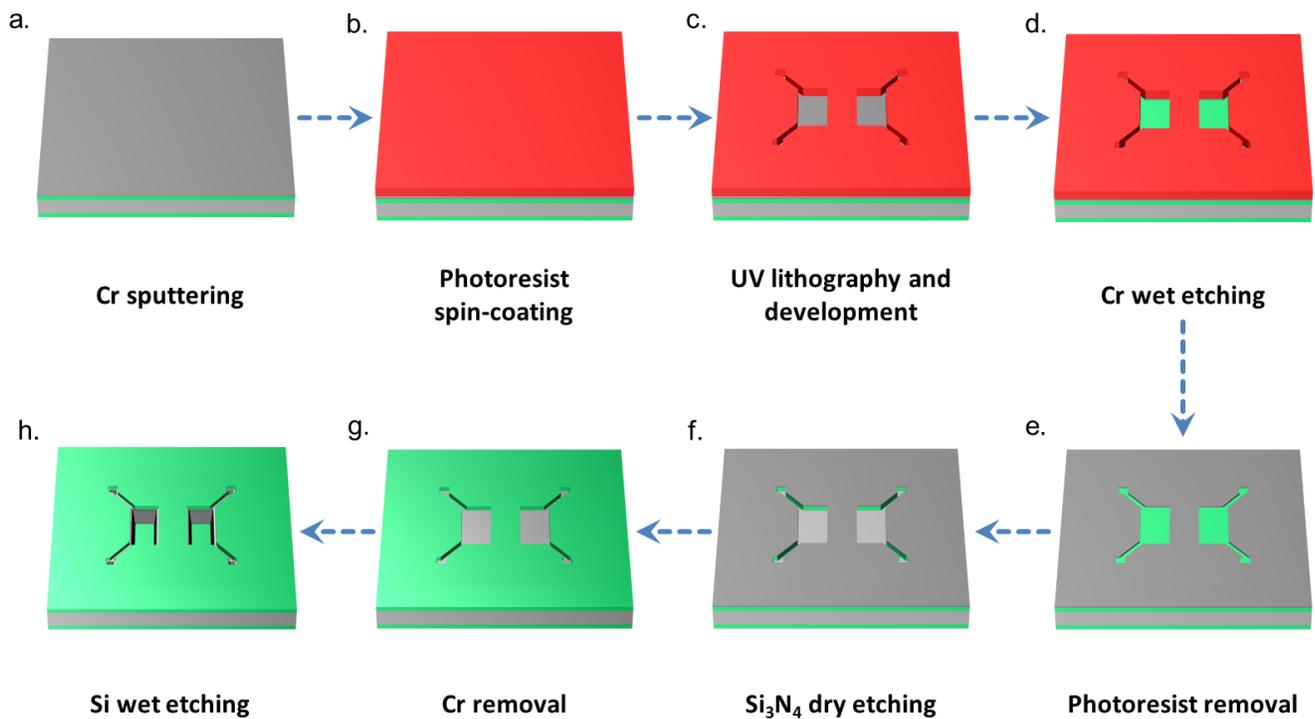

**Figure S1**. Schematic representation of membranes and micro-channels fabrication protocol.

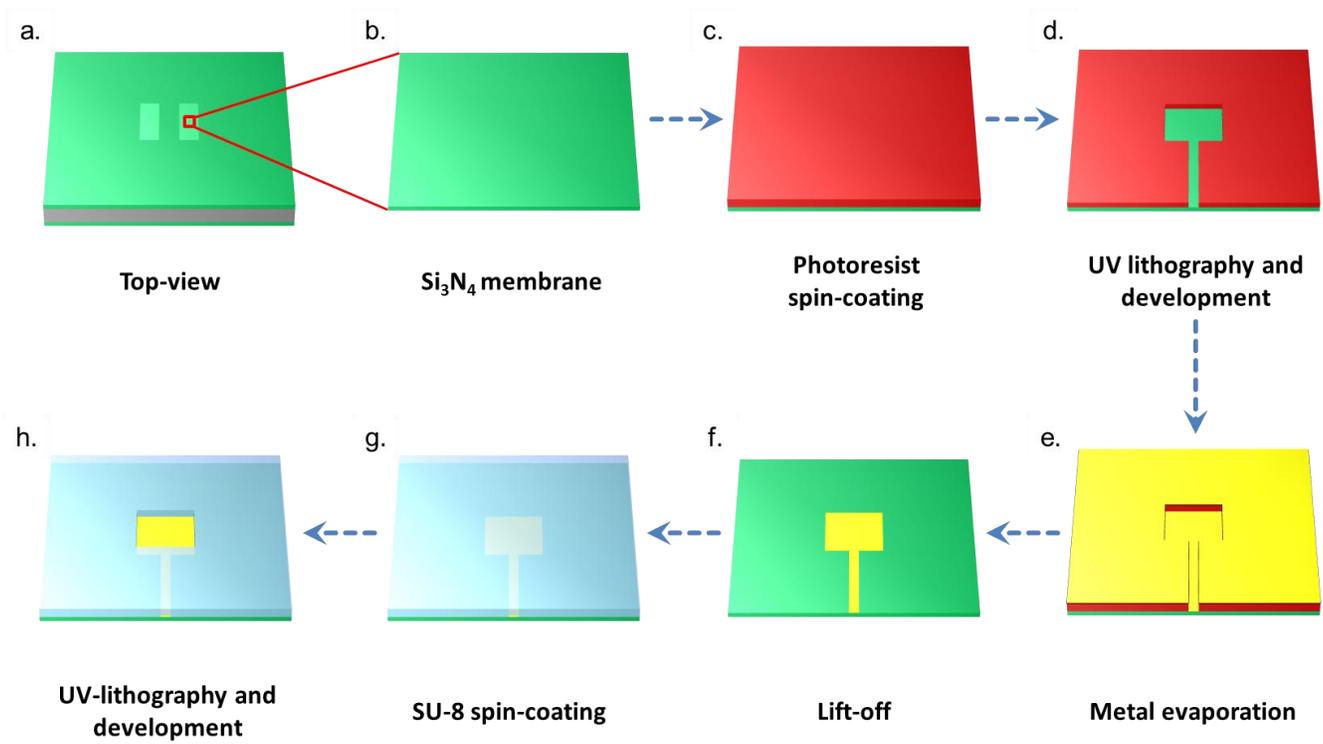

**Figure S2**. Schematic representation of the microfluidic MEA fabrication protocol.

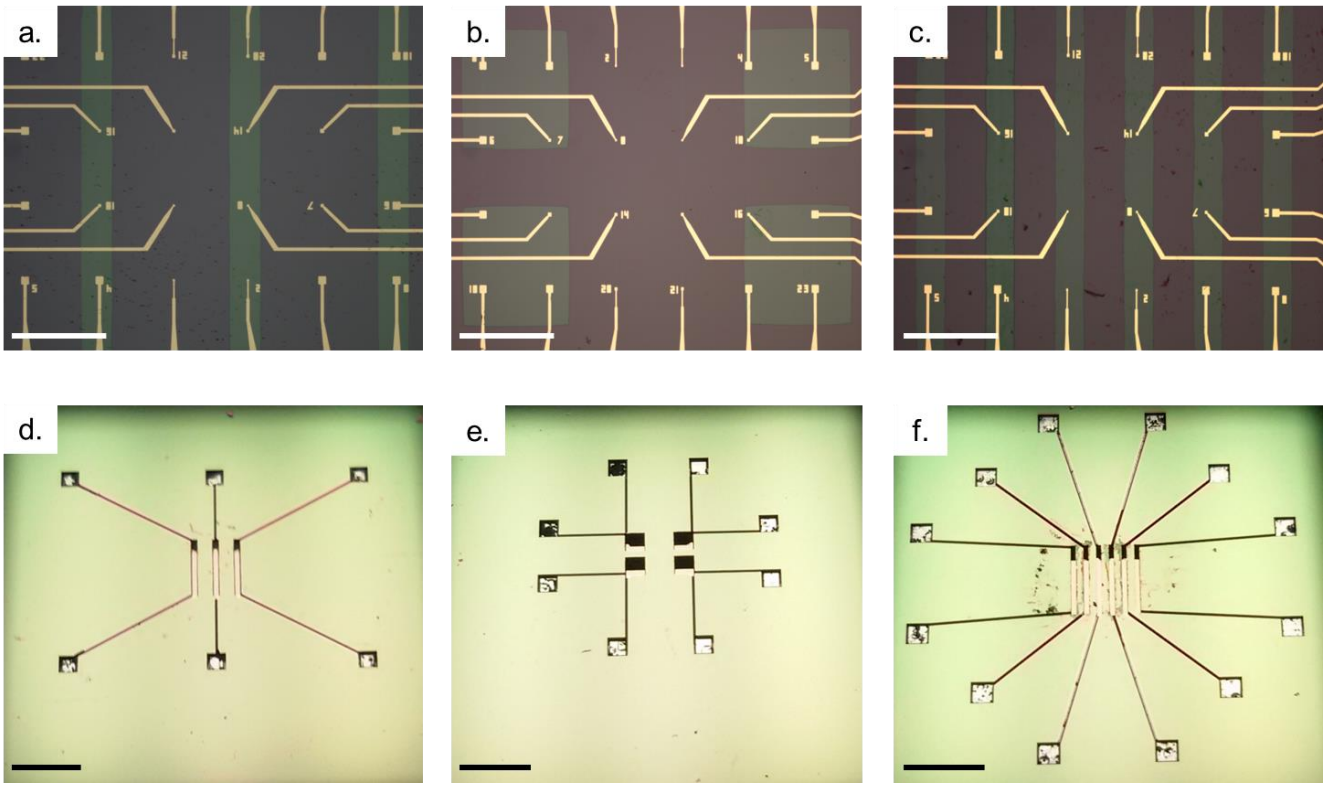

**Figure S3**. Different microfluidic arrangements. **a**)-**c**) Optical images of different microfluidic configuration, respectively with 3, 4 and 6 micro-channels, acquired from the recording side of the device. The scale bar is 500 μm. **d**)-**f**) Optical images of different microfluidic configuration, respectively with 3, 4 and 6 micro-channels, acquired from the delivery side of the device. The scale bar is 2.5 mm.

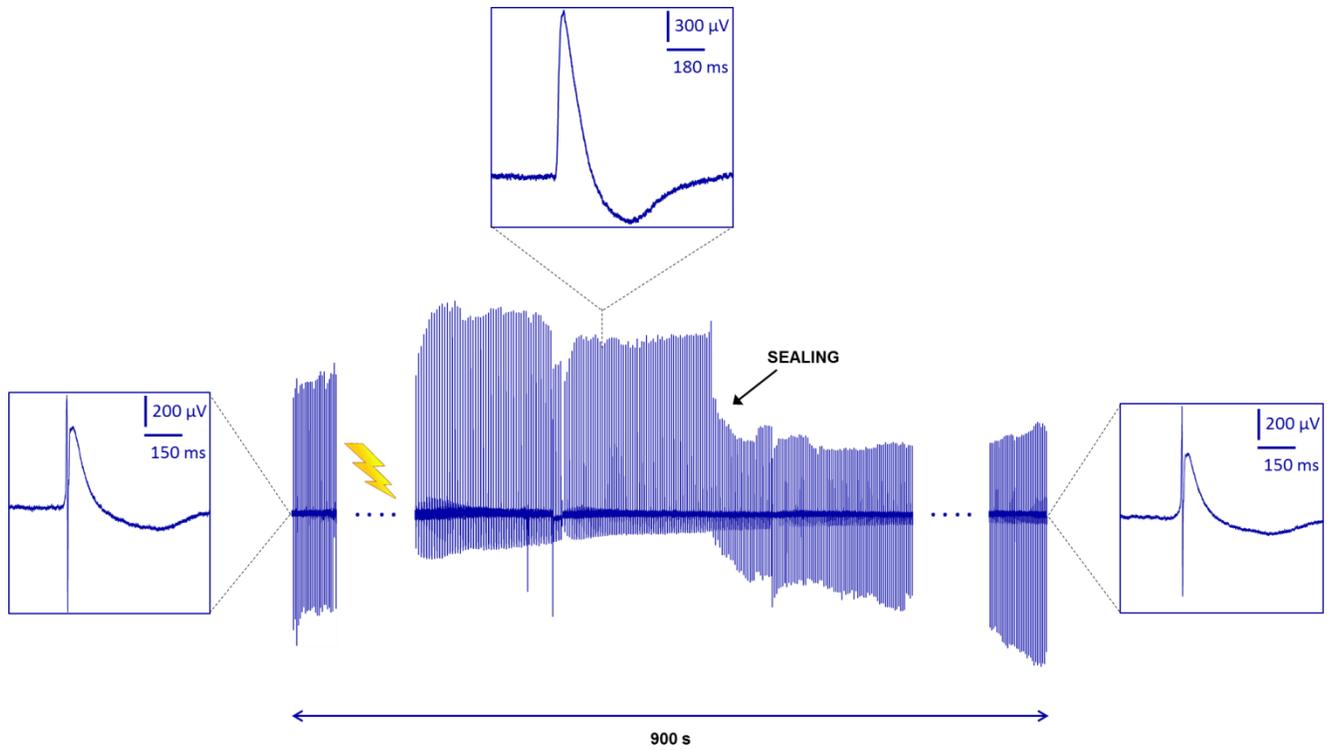

**Figure S4**. Long recording on hiPSC cardiomyocytes. After electroporation, the spontaneous electrophysiological spiking activity is modified into intracellular-like spikes. The spontaneous activity resumes extracellular features after 10 – 15 minutes upon resealing of the transient membrane pores.

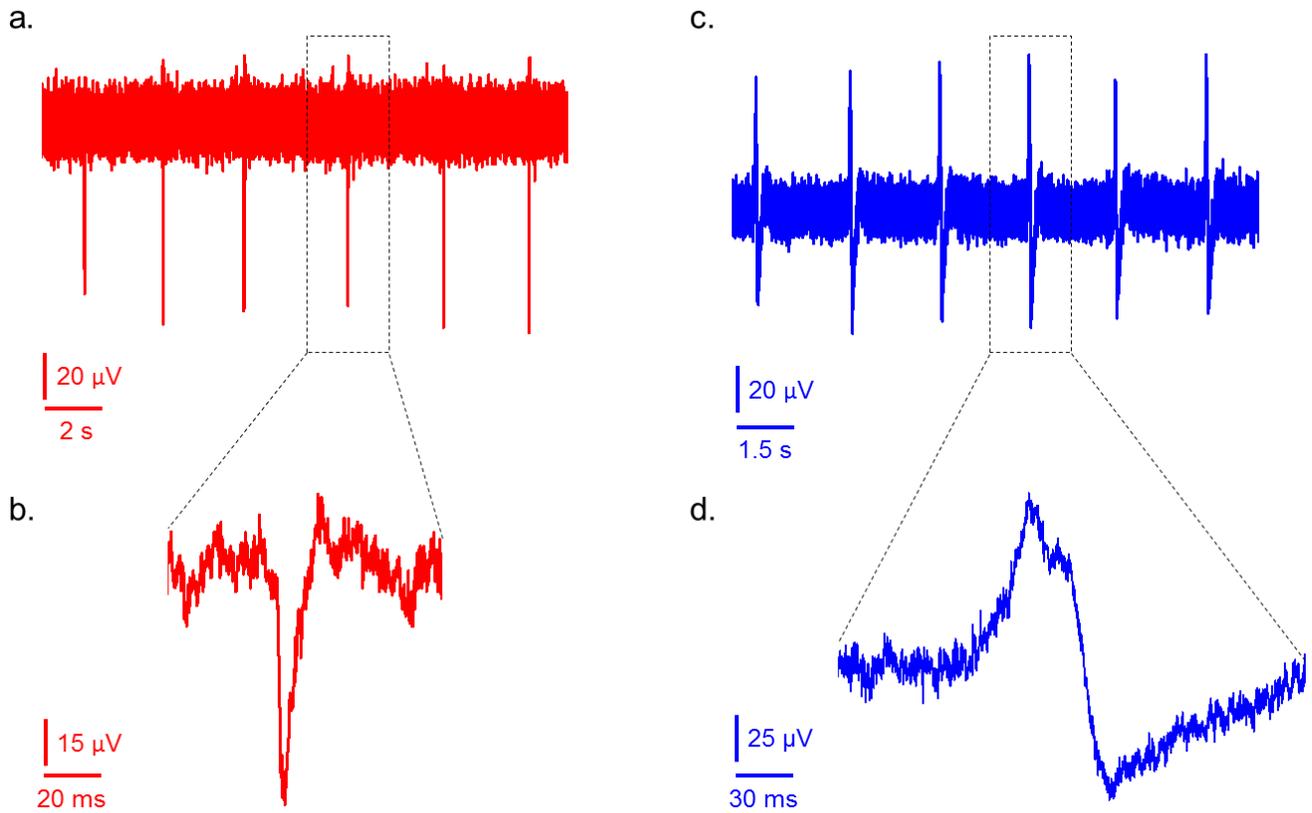

**Figure S5**. Extra and intra-cellular recording on HL-1 cardiomyocytes. **a)-b**) Spontaneous electrophysiological spiking activity of HL-1 cardiomyocytes from one of the 24 channels of a MF-MEA. **c)-d**) Intracellular-like activity of HL-1 cardiomyocytes acquired on the same electrode after electroporation.

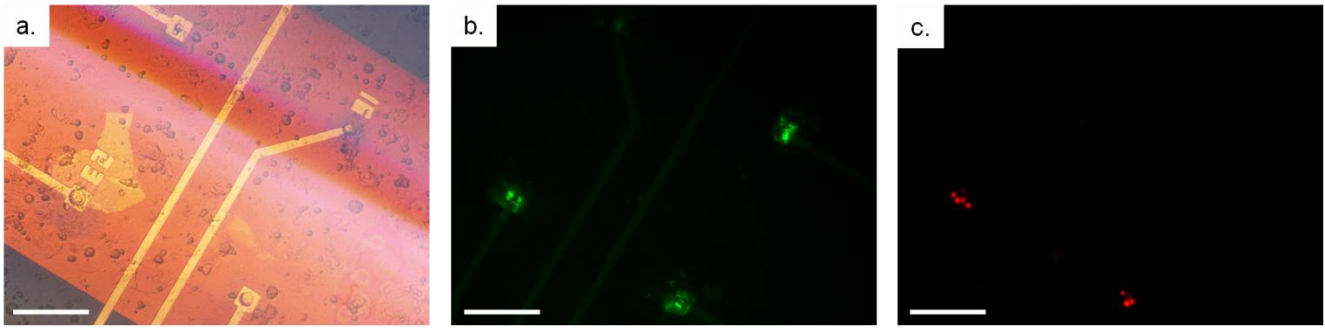

**Figure S6.** Selective intracellular drug delivery on MF-MEA. **a**) Bright-field optical image of four electrodes of a MF-MEA. The scale bar is 200 μm. **b**) The same 4 electrodes observed with a green fluorescence filter, showing calcein-AM delivery to the cells onto the 4 electrodes. The scale bar is 200 μm. **c**) The same four electrodes observed with a red fluorescence filter, showing PrhD-1 delivery to the cells onto the 2 electrodes where electroporation was performed. The scale bar is 200 μm.

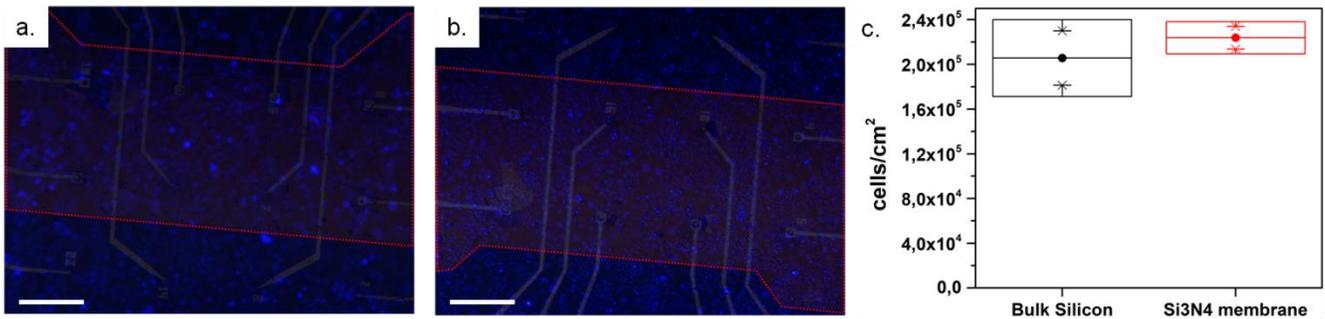

**Figure S7.** Cell density estimate after 4 DIV. **a**) Fluorescence image of cardiomyocytes cultured on MF-MEA and stained with DAPI after 4 DIV. The red-bordered area delimits the nitride membrane. The scale bar is 300 μm. **b**) Fluorescence image, acquired on the other membrane region of the same device, of cardiomyocytes cultured on MF-MEA and stained with DAPI after 4 DIV. The red-bordered area delimits the nitride membrane. The scale bar is 300 μm. **c**) Box plot featuring the cell density distribution calculated for different areas (bulk silicon or nitride membranes) on the same device.